# Structural, Magnetic and Magneto-caloric studies of $Ni_{50}Mn_{30}Sn_{20}$Shape Memory Alloy

Ramesh Chandra Bhatt, R.S. Meena, H. Kishan, V.P.S. Awana* and S.K. Agarwal
Quantum Phenomena and Applications Division, CSIR-National Physical Laboratory, New Delhi-110012, India

## Abstract

We have synthesized a nominal composition of $Ni_{50}Mn_{30}Sn_{20}$ alloy using arc melting technique. Rietveld refinement confirms the austenite $L2_1$ structure in *Fm-3m* space group. Electrical resistivity has been found to clearly exhibiting two different phenomena viz. a magnetic transition from paramagnetic to ferromagnetic and a structural transition from austenite to martensitic phase. Thermo-magnetization measurements M(T) confirms ferromagnetic transition temperature $T_C$ at 222 K and martensitic transition starting at 127 K($M_S$). Magnetization measurement M(H) at 10 K confirms the ferromagnetic state. Frequency dependence of ac susceptibility $\chi'$ at low temperature suggests spin glass behavior in the system. The isothermal magnetic entropy change values have been found to be 1.14 J/Kg.K, 2.69 J/Kg.K and 3.9 J/Kg.K, with refrigeration capacities of 19.6 J/kg, 37.8 J/kg and 54.6 J/kg for the field change of 1, 2 and 3 Tesla respectively at 227 K.

****Corresponding Author***awana@nplindia.org
Tel.: +91-11-45609357; Fax: +91-11-45609310
***Web page-*** awanavps.webs.com

## 1. **Introduction**

In the recent past intermetallic compounds have attracted considerable attention for their abilities to show interesting structural, magnetic and superconducting (metal pnictide, chalcogenide FeSe/As) properties[1]. The existence of complex meta-magnetic transition in these compounds makes them potential candidates for magnetic studies. Their magnetic behavior is still uncertain as some of these show signatures of spin glass in the low temperature regime[2]. Due to their intriguing magnetic properties researchers have investigated these materials for magneto-resistance and magneto-caloric aspects[3–7].Further, excellent shape memory effect has also been demonstrated with alloys like NiTi (nitinol) [8,9]corresponding to martensite - austenite phase transition which are being widely used in military, medical and robotic applications. Due to the presence of meta-magnetic behavior there are possibilities for these materials to show normal as well as inverse magneto-caloric effects. The magneto-caloric effect is the change in the magnetic entropy or lattice entropy when subjected to an external magnetic field isothermally or adiabatically[10]. In general, near ferromagnetic transition material exhibit normal magneto-caloric effect, on the other hand inverse magneto-caloric effect is observed for antiferromagnetic transition[10]. Comparatively easy synthesis process of the intermetallic compounds with large magnetic moment values have generated possibilities of a cheaper, efficient and environment friendly magnetic refrigerator. In NiMn-family, Ni-Mn-Ga ferromagnetic shape memory alloys have been extensively studied for magnetic cooling applications[11–14]. But their practical usage have been limited due to the operating temperature range and high cost of Ga. Ga has been replaced by In and Sn. Sutou et al.[15] studied $Ni_{50}Mn_{50-x}Sn_x$ ferromagnetic Heusler alloys and found martensitic transformation in the system. Krenke et al.[1] reported inverse magneto-caloric effect in NiMnSn alloys and found sharp peak of magnetic entropy change corresponding to a sharp martensitic transition whereas Krenke et al.[16] investigated the martensitic transition in such Heusler alloys. Sharma et al.[17] studied $Ni_{50}Mn_{34}Sn_{16}$ alloy for the magneto-caloric effect and found conventional (normal) and inverse magneto-caloric effect for para-ferromagnetic transition and martensitic transition respectively. Llamazares et al.[18] studied inter-martensitic phase transition in $Ni_{50}Mn_{38}Sn_{12}$ alloy ribbons. Choudhary et al.[19] studied mechanical and damping properties of room temperature ferromagnetic $Ni_{50.4}Mn_{34.7}Sn_{14.9}$:Ti composite thin films and found increased martensitic transformation temperature with Ti content. Ghosh et al.[20] studied magneto-caloric properties of NiMnSn alloys and found increased value of magnetic entropy change with martensitic transformation shifting towards high temperature with Mn content. There are only a few reports about the reentrant spin glass behavior on Ni-Mn-Sn system. Chatterjee et al.[2] have observed reentrant spin glass state in $Ni_2Mn_{1.36}Sn_{0.64}$ shape memory alloys. The diversity in the magnetic and magneto-caloric behavior in the Ni-Mn-Sn system has motivated us to investigate this system. In the present communication we have synthesized and investigated structural, magnetic and magneto-caloric behavior of $Ni_{50}Mn_{30}Sn_{20}$ Heusler alloy which is seen to exhibit signature of spin glass in the system.

## 2. Experimental

A button ingot with nominal composition of $Ni_{50}Mn_{30}Sn_{20}$ (atomic %) has been synthesized by arc melting nickel, manganese and tin. These metals were mixed and pelletized in a pelletizer and then melted several times using single arc in argon gas atmosphere. Such as-prepared alloy was sealed in an evacuated quartz tube for annealing at temperature $1000^0$C with a heating rate of $120^0$C/hour for 24 hours and quenched thereafter in ice water. To examine its phase purity and determining lattice parameters, room temperature X-ray diffraction measurement has been performed through powder diffractometer (Rigaku Miniflex II) using $CuK_\alpha$ radiation. Reitveld refinement of XRD pattern has been performed using Fullprof programme. Electrical transport and magnetic measurements have been performed on a physical property measurement system (PPMS, 14 Tesla, Quantum design make). The magnetization measurements with temperature M(T) have been performed in ZFC, FC and FCW configurations. In ZFC mode the sample was cooled in the absence of magnetic field and the magnetization data was recorded during the warming. In FC mode sample was cooled in the presence of magnetic field and the data was measured during cooling while in FCW mode data was recorded during warming. The AC susceptibility measurements have been performed in different frequencies and amplitudes using PPMS. The isothermal magnetization measurements at discrete temperature intervals have also been performed for the estimation of magnetic entropy change.

## 3. Results and discussion

Rietveld refinement of the X-ray diffraction (XRD) pattern confirms the presence of all the identified characteristic peaks exist in $L2_1$cubic crystal structure with *Fm-3m* space group (# 225) listed in fig. 1. The blue line in fig. 1 shows the difference in the observed and fitted pattern. The (hkl) listed XRD pattern also confirms the austenite phase at room temperature. The standard Wyckoff positions being used for Rietveld refinements of $Ni_{50}Mn_{30}Sn_{20}$ are given in table 1. The quality of fit parameter ($\chi^2$) obtained after refinement was 7.82 with $R_{wp}$= 31.7.The unit cell corresponding to the refined cell parameter value 6.037 Å has been shown in fig. 2.

Figure 3shows the temperature dependence of normalized resistivity $\rho/\rho_{300K}(T)$ for the $Ni_{50}Mn_{30}Sn_{20}$ alloy. The resistivity data has been recorded while cooling the sample. By observing resistivity data in these systems one can detect structural as well as magnetic transitions. The overall behavior of resistivity is metal-like except for a change in slope at high temperature and a resistivity jump at lower temperature near 130 K. At high temperatures the change in the slope corresponds to the magnetic transition i.e. paramagnetic

- ferromagnetic transition marked as $T_C$ in the curve. The resistivity jump near 130 K corresponds to the martensitic transformation with $M_S$ and $M_F$ as martensitic start and finish temperatures respectively. Similar kind of behavior has also been observed by Vasil'ev at al.[21]in $Ni_{2+x}Mn_{1-x}Ga$ shape memory alloy.

Figure 4 shows thermo-magnetization curve obtained during cooling and warming mode at applied magnetic field of 100 Oe. The sample shows paramagnetic– ferromagnetic transition with $T_C$ at ~ 222 K in the austenitic phase. The observed $T_C$ in our sample is less than the similar stoichiometric alloys(i.e. $Ni_{50}Mn_{50-x}Sn_x$)[15–18,20,22] which is probably ascribed to deficiency of Sn in the system. In martensitic transformation region(127K –80K) of the system under study, the coexistence of antiferromagnetic and ferromagnetic exchange interactions has been observed. The possible deficiency of Sn creates off-stoichiometry in the system and the Mn atoms may occupy the vacant Sn sites. The antiferromagnetic behavior in this system is due to the antiferromagnetic coupling of Mn atoms with the excess Mn atoms (off-stoichiometric) situated at the Ni/Sn sites[18]. On warming the austenite phase is found to reoccur showing the shape memory effect. The $A_S$ and $A_F$ have been marked in fig. 4as the start and finish points of austenitic transformation. In martensitic phase we have observed a magnetic moment of 3.5emu/g for 100Oe applied field. Figure 5 shows the magnetization M(H) curve at 10 K. It is clear from the M(H) curve that in the martensitic phase the starting alloy $Ni_{50}Mn_{30}Sn_{20}$exhibits a clear ferromagnetic nature. It also suggests that martensitic transformation has not completely destroyed the ferromagnetism.

AC susceptibility measurements have been performed at different frequencies to investigate the possibility of existence of any spin glass behavior. Figure 6 shows the real part of ac susceptibility measurements $\chi'(T)$ at applied field amplitudes of 3, 5 and 10 Oe with frequencies 333, 666, 999, 3333, 6666 and 9999 Hz. We observed a paramagnetic–ferromagnetic transition followed by martensitic transformation which is also supported by the dc thermo-magnetization measurement (fig.4). At low temperatures a shift in susceptibility χ' value with frequency for 10 Oe applied field has been observed and is depicted in the inset of fig. 6. It is evident from the inset that the patterns shift to higher temperature range with increase in frequency, suggesting spin glass behavior in this region[23]. In the present case it is conjectured that the martensitic transition and the antiferromagnetic coupling of Mn atoms reduce the spin alignment, and the competition among these phenomena gives rise to spin frustration in the system which apparently could be source of the spin glass behavior observed in the system. Evidence of spin glass behavior in these systems has been reported earlier also[24].

The isothermal magnetic entropy change ($\Delta S_M$) values at different applied magnetic fields have been estimated by the following equation using classical Maxwell's thermodynamic relation [25];

$$\Delta S_M\,(T,H) = \int_0^{H_m} \left(\frac{\partial M}{\partial T}\right)_H dH$$

When magnetization isotherms performed at small discrete field temperature intervals, this integral can be approximated as below [26]

$$\lfloor \Delta S_M \rfloor = \sum_i \frac{M_i + M_{i+1}}{T_{i+1} + T_i} \Delta T H_i$$

Here, $M_i$ and $M_{i+1}$ are the experimental values of magnetization at temperature $T_i$ and $T_{i+1}$ respectively. The $\Delta S_M$ values at different temperatures have been calculated using the magnetization isotherms taken in the temperature range 180 to 250 K at the interval of 5 and 10 K around $T_C$ as depicted in fig. 7.In the present study we were aiming to achieve magnetic entropy change near room temperature therefore we have performed magnetization around $T_C$ for conventional MCE and not in the martensitic regime. Temperature dependence of the magnetic entropy change for $Ni_{50}Mn_{30}Sn_{20}$has been shown in fig. 8.The obtained values of|$\Delta S_M$| (modulus for negative values) are 1.14 J/Kg.K, 2.69 J/Kg.K and 3.9 J/Kg.K for the field change of 1Tesla, 2Tesla and 3Tesla respectively at 227 K (table 2). We observed magnetic entropy change values to be spread around a wide temperature range which is significant requirement for an active magnetic refrigerant material. Therefore we have estimated refrigeration capacity (RC) from the calculated $\Delta S_M$ values. RC of the material has been estimated by taking the product of peak |$\Delta S_M$| value and full-width at half maxima ($\delta T_{FWHM} = T_2$-$T_1$) in the temperature dependent magnetic entropy change plot.

$$RC= |\Delta S_M|\times\delta T_{FWHM}$$

The calculated RC values have been tabulated in table2. The $\Delta S_M$ and RC values achieved in the present study found to be comparable with the existing values in the literature for the similar Ni-Mn-Sn system. Ghosh et al. reported a maximum value of |$\Delta S_M$|~ 5.15 J/Kg.K and RC value 35.5 J/kg at 1.5 Tesla field in $Ni_{50}Mn_{36.5}Sn_{13.5}$[27]and RC 36.48 J/kg at 1.5 Tesla field in $Ni_{46.5}Co_1Mn_{37.5}Sn_{15}$[5]. Sharma et al. reported|$\Delta S_M$| value of 4 J/Kg.K at 8 Tesla field for$Ni_{50}Mn_{34}Sn_{16}$ [17]. Han et al. reported sharp |$\Delta S_M$| peak of 10.4 J/Kg.K at 1 Tesla field for $Ni_{43}Mn_{46}Sn_{11}$[28].

## Conclusions

We have synthesized a nominal composition of $Ni_{50}Mn_{30}Sn_{20}$ alloy using arc melter. Rietveld refinement confirms the austenite $L2_1$structure in *Fm- 3m*space group. Electrical resistivity clearly shows the two different phenomena viz., magnetic transition from paramagnetic to ferromagnetic and the structural transition from austenite to martensite phase. Thermo-magnetization measurements confirm these two phenomena and magnetization

M(H)confirms ferromagnetic state at 10 K. The frequency dependence of ac susceptibility at low temperature suggests possible spin glass behavior in the system. The magnetic entropy change $|\Delta S_M|$ has been calculated from magnetization isotherms around $T_C$ and shows reasonable values in a wide temperature range. The $|\Delta S_M|$ and refrigeration capacities (RC) of the prepared alloy have been found to be comparable with their respective values on similar NiMnSn system.

**Acknowledgements**

Authors thank the Director, National Physical Laboratory, New Delhi, for his support.

**Table and Figure Captions:**

**Table 1.** Wyckoff positions for $Ni_{50}Mn_{30}Sn_{20}$Heusler alloy in space group:*Fm-3m*(# 225).

**Table 2.** Experimental values of magnetic entropy change $|\Delta S_M|$ and refrigeration capacity RC of $Ni_{50}Mn_{30}Sn_{20}$.

**Figure 1.** Rietveld refined XRD pattern of$Ni_{50}Mn_{30}Sn_{20}$ alloy.

**Figure 2.** Cubic unit cell corresponding to Rietveld refined XRD patterns of $Ni_{50}Mn_{30}Sn_{20}$ alloy.

**Figure 3.** Temperature dependent electrical resistivity $\rho/\rho_{300K}(T)$of $Ni_{50}Mn_{30}Sn_{20}$ alloy.

**Figure 4.** DC thermo-magnetization of $Ni_{50}Mn_{30}Sn_{20}$ alloy in field cooled (FC) and field cooled warming (FCW) modes.

**Figure 5.** Magnetization M(H) hysteresis loop of studied $Ni_{50}Mn_{30}Sn_{20}$ alloy at 10K.

**Figure 6.** Temperature dependence of real part of ac susceptibility $\chi'(T)$ at different field amplitudes and frequencies, showing ferromagnetic, martensitic and spin glass behavior. The inset shows enlarged view of $\chi'$ shift with frequency at lower temperature range.

**Figure 7.** Magnetization isotherms of studied $Ni_{50}Mn_{30}Sn_{20}$ alloy in 180-250K temperature range.

**Figure 8.** Magnetic entropy change values ($\Delta S_M$) around $T_C$ at field change of 1, 2 and 3 Tesla for the studied $Ni_{50}Mn_{30}Sn_{20}$ alloy.

**Table 1**

| *Atom* | *Site* | *x* | *y* | *z* |
|---|---|---|---|---|
| *Ni* | *8c* | *1/4* | *1/4* | *1/4* |
| *Mn* | *4a* | *0* | *0* | *0* |
| *Sn* | *4b* | *1/2* | *1/2* | *1/2* |

**Table 2**

| Sample | $\|\Delta S_M\|$ (J/Kg.K) | | | RC (J/Kg) | | |
|---|---|---|---|---|---|---|
| | 1Tesla | 2Tesla | 3Tesla | 1Tesla | 2Tesla | 3Tesla |
| $Ni_{50}Mn_{30}Sn_{20}$ | 1.41 | 2.69 | 3.9 | 19.25 | 38.39 | 54.6 |

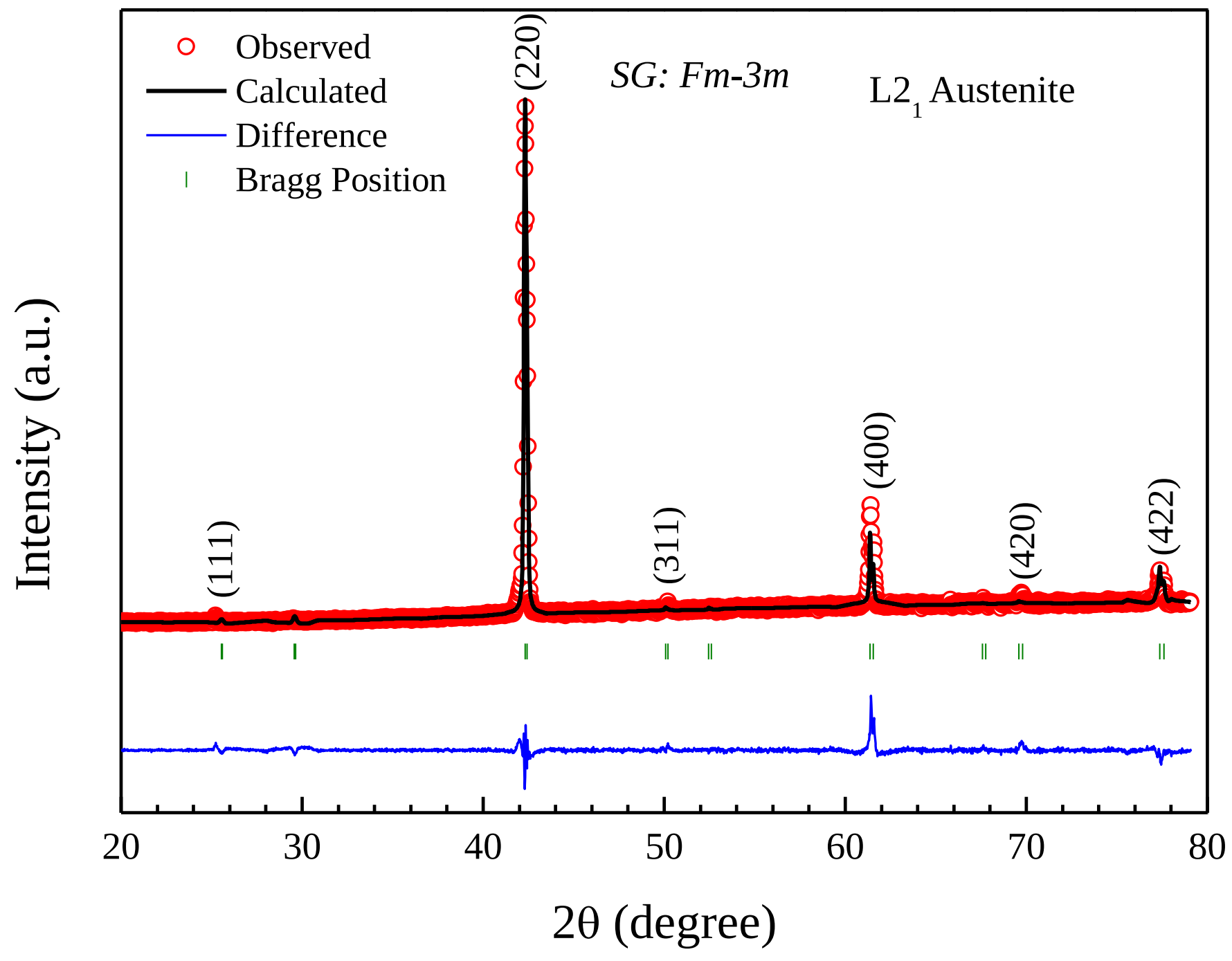

**Figure 1**

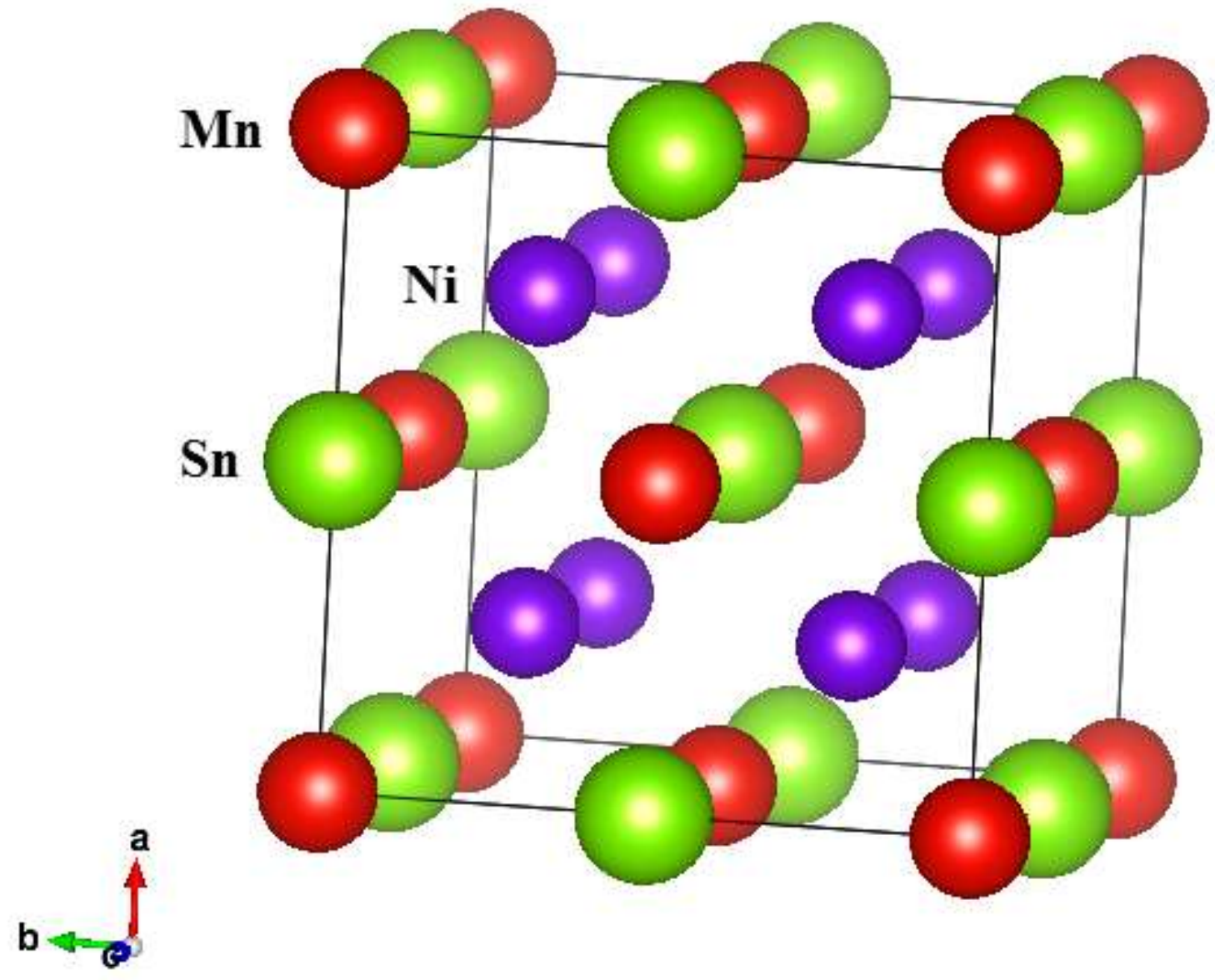

**Figure 2**

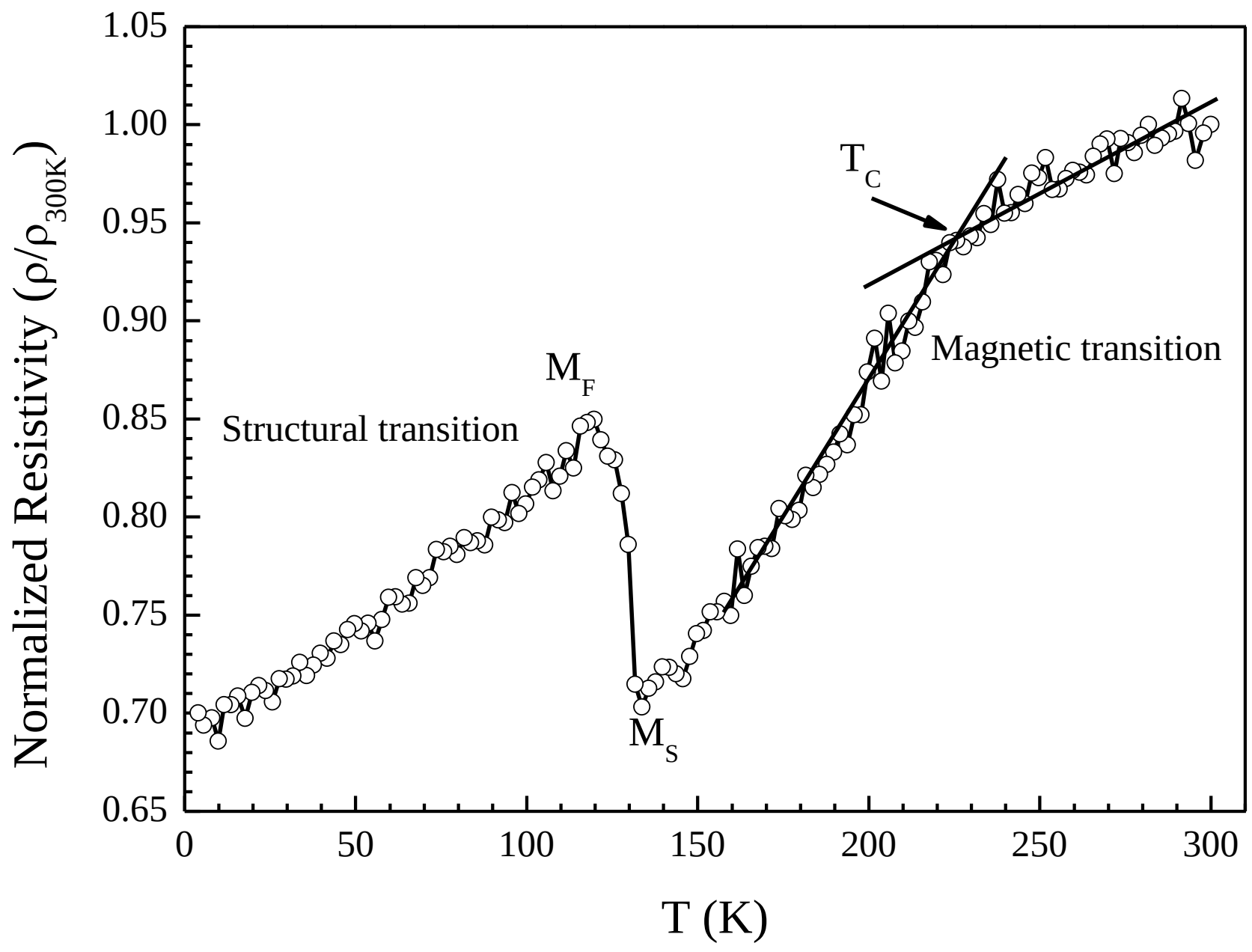

**Figure3**

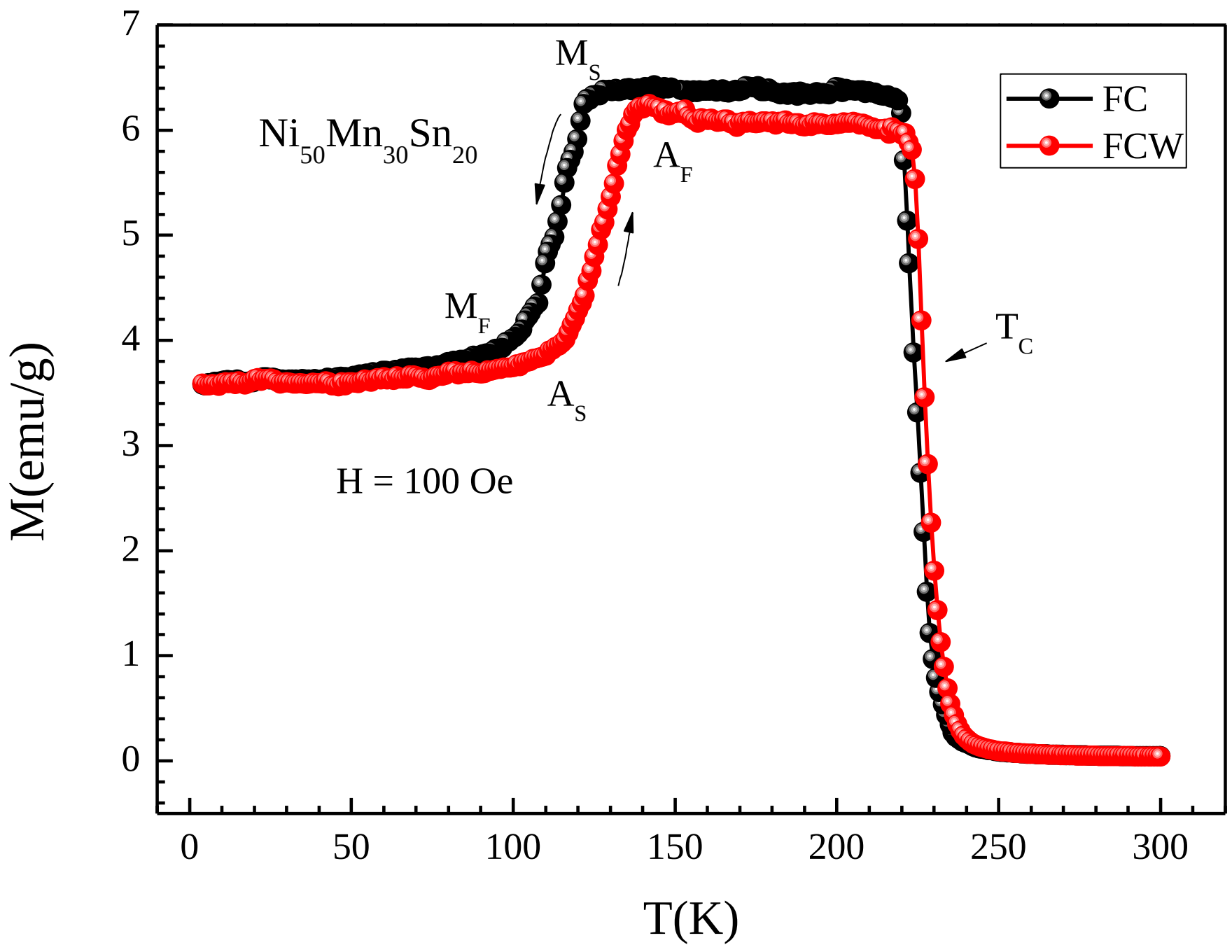

**Figure 4**

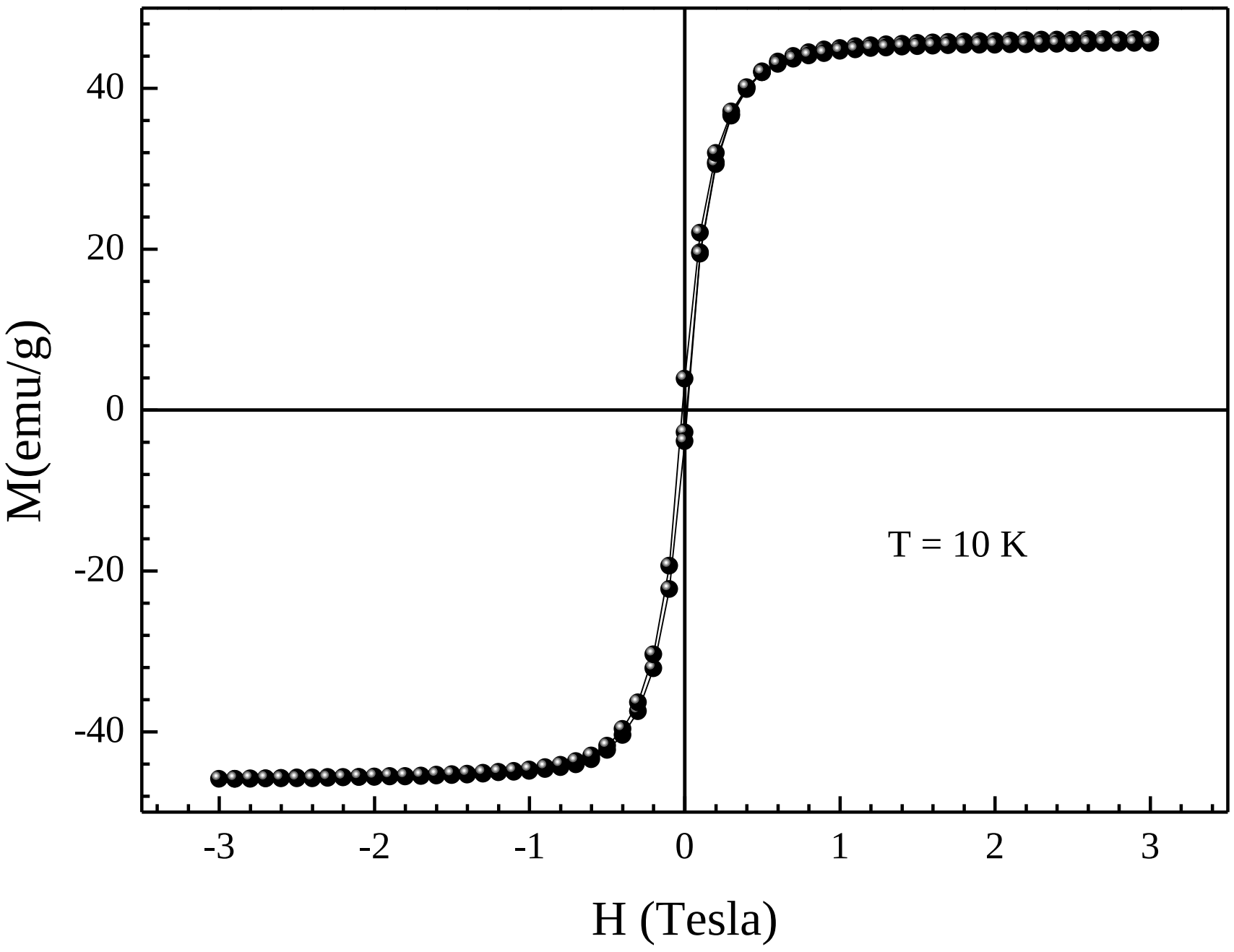

**Figure 5**

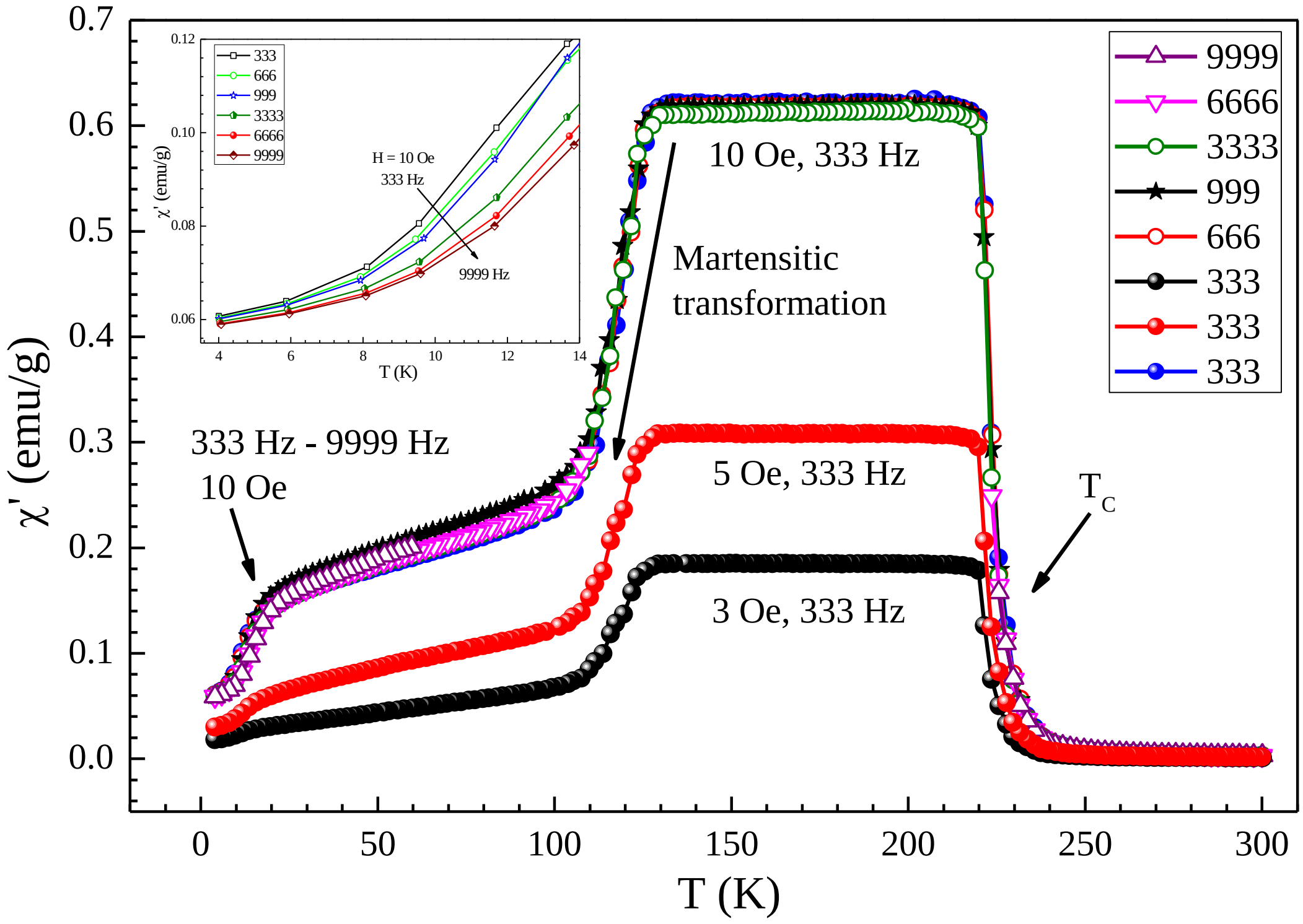

**Figure 6**

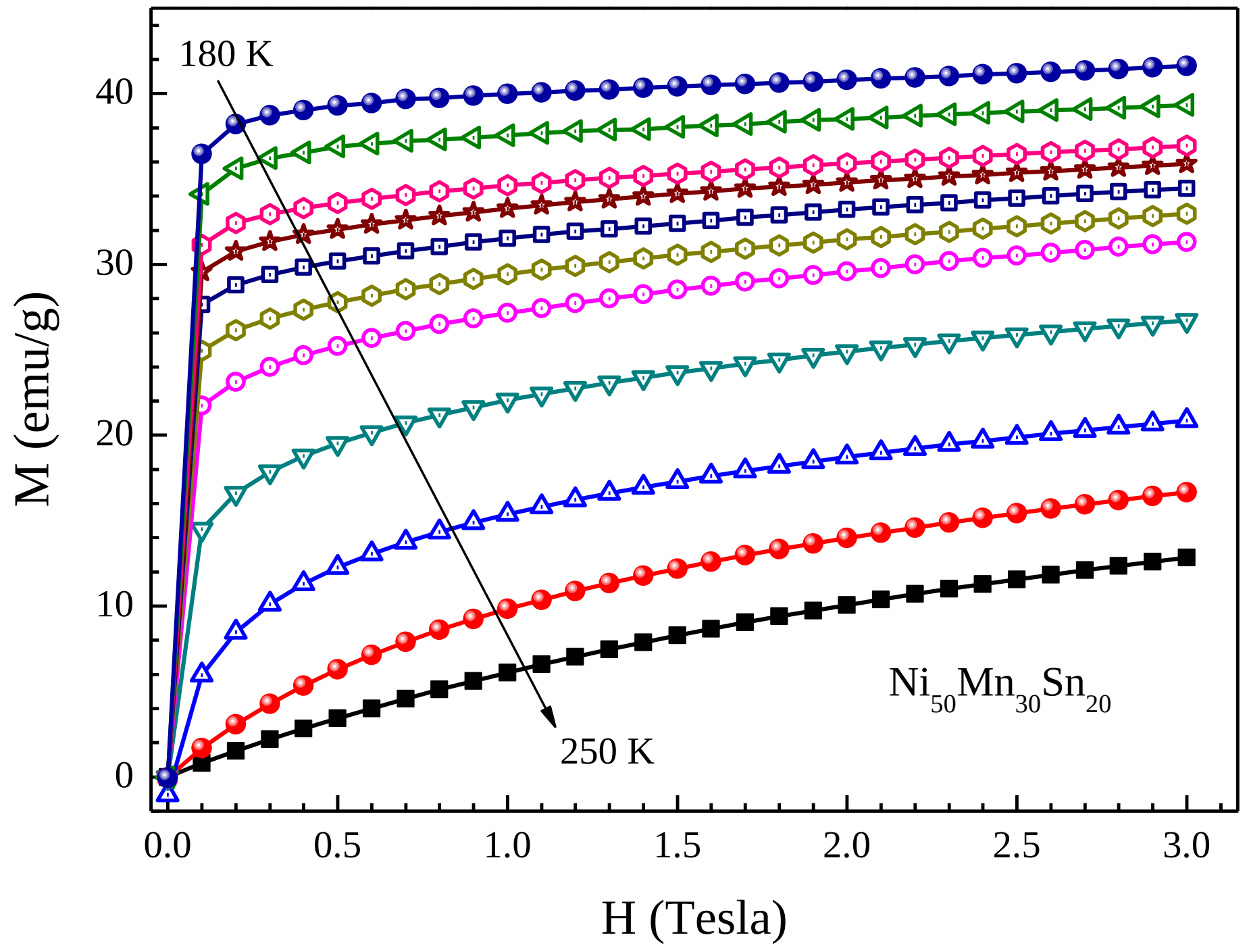

**Figure 7**

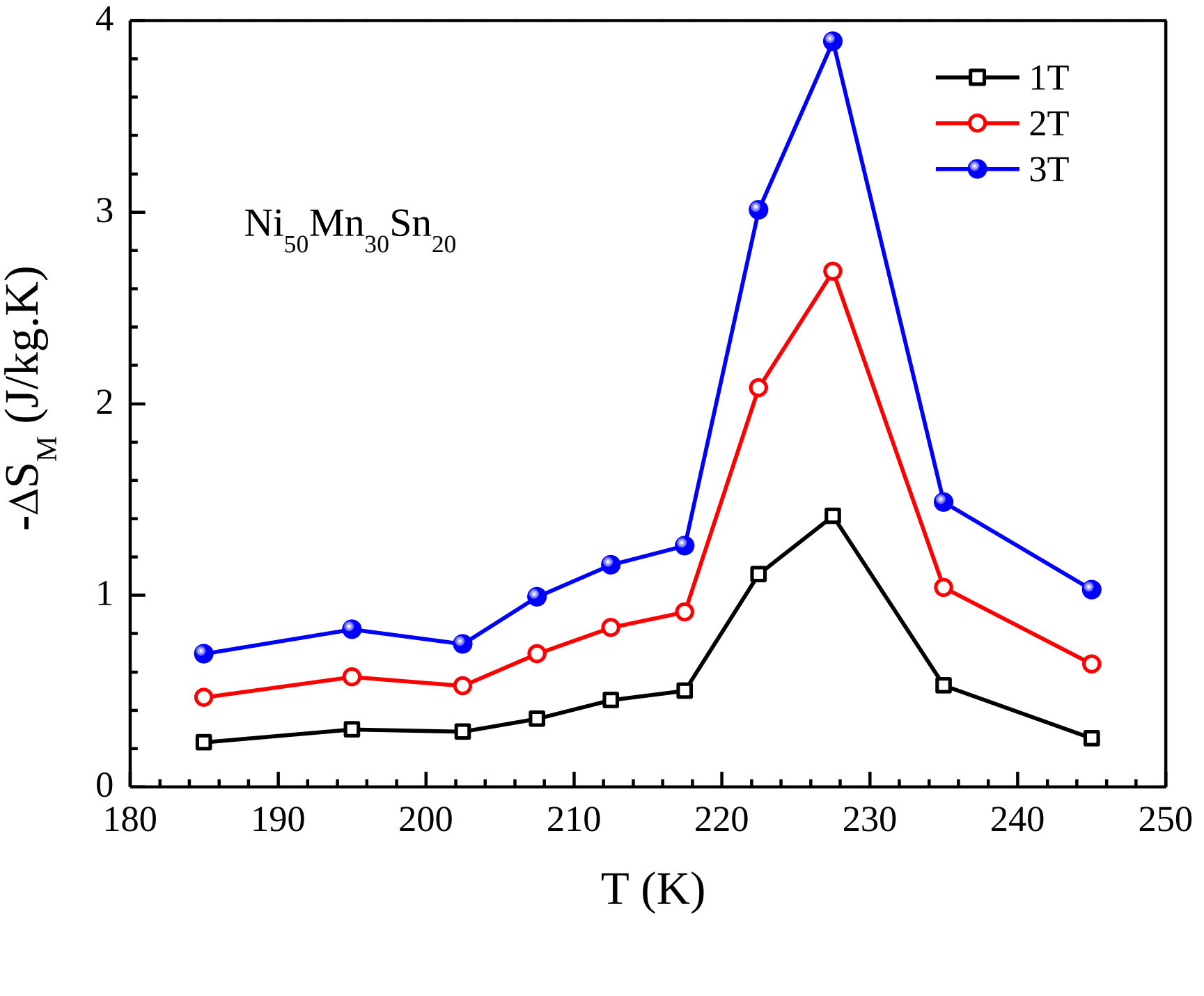

**Figure8**